\documentclass[prl,twocolumn,showpacs]{revtex4}
\usepackage{epsfig}
\usepackage{graphicx}
\begin{document}

\title{Fermions in 2D Optical Lattices: Temperature and
Entropy Scales for Observing Antiferromagnetism and Superfluidity}
\author{Thereza Paiva$^{1}$, Richard Scalettar$^{2}$, Mohit Randeria$^{3}$, and Nandini Trivedi$^{3}$}
\affiliation{$^{1}$Instituto de Fisica, Universidade Federal do Rio de Janeiro Cx.P. 68.528, 21945-970 Rio de Janeiro RJ, Brazil\\
$^{2}$Department of Physics, University of California, Davis, CA 95616, USA\\
$^{3}$Department of Physics, The Ohio State University, Columbus, OH 43210, USA }

\begin{abstract}
One of the major challenges in realizing antiferromagnetic and
superfluid phases in optical lattices is the ability to cool fermions.
We determine constraints on the entropy for observing these phases in
two-dimensional Hubbard models. We investigate antiferromagnetic
correlations in the repulsive model at half filling and superfluidity of
s-wave pairs in the attractive case away from half filling using
determinantal quantum Monte Carlo simulations that are free of the
fermion sign problem.  We find that an entropy per
particle $\simeq \ln 2$ is sufficient to observe the
charge gap in the repulsive Hubbard model or the pairing pseudogap in
the attractive case. Observing antiferromagnetic correlations or
superfluidity in 2D systems requires a further reduction in
entropy by a factor of three or more.  In contrast to higher dimensions, we find that
adiabatic cooling is not useful to achieve the required low
temperatures. We also show that double occupancy measurements are useful
for thermometry for temperatures greater than the nearest-neighbor
hopping. 

\end{abstract}

%    74.72.-h, 74.20.-z, 75.10.Jm, 71.27.+a \textcolor{red}{CHECK PACS!}}
% 71.10.Fd	Lattice fermion models (Hubbard model, etc.)
%71.27.+a	Strongly correlated electron systems; heavy fermions
%37.10.Jk	Atoms in optical lattices
\maketitle

An exciting new development in cold atoms is the ability to realize in
the laboratory simple models of strongly correlated fermions in optical
lattices \cite{esslinger,ketterle,jordens,schneider,hofstetter,zoller}.
These studies are motivated by their relevance to spectacular phenomena
in condensed matter physics, like high $T_c$ superconductivity, that
are not fully understood.  The best known model is the fermion Hubbard
Hamiltonian\cite{scalapino,andersonAtoZ} that captures the
physics of antiferromagnetism and, at least qualitatively, 
d-wave superconductivity in two dimensions (2D). The Hubbard model is
well understood in one dimension (1D), using exact solutions and
bosonization \cite{giamarchi}, and also in the limit of large
dimensions, using dynamical mean field theory (DMFT) \cite{georges}. The
\emph{two dimensional} problem, of direct relevance to layered high $T_c$
superconductors, is the least well understood theoretically. 
New insights into the 2D Hubbard model can come from cold atom emulators,
given their high degree of tunability (interaction strength, chemical
potential) and absence of disorder and material complications.

The principal challenges for the optical lattice emulators are to cool
down to sufficiently low temperature to see interesting phases and to
observe the characteristic order. In this paper we present detailed
quantitative results from quantum Monte Carlo (QMC) simulations
\cite{scalapino} of (i) the 2D repulsive ($U>0$) Hubbard model at half
filling and (ii) the 2D attractive ($U<0$) Hubbard model at any filling,
that are both of direct relevance to ongoing experiments.  The reasons
for focusing on these systems are threefold.  Determinantal QMC
simulations are free of the fermion ``sign problem'' \cite{hirsch} in
both cases and we can obtain non-perturbative results on finite systems
without any approximations. These two systems exhibit phenomena of
great interest: strong antiferromagnetic correlations and Mott physics
for $U>0$, and a Berezinskii-Kosterlitz-Thouless (BKT) transition to an s-wave superfluid
\cite{moreo-scalapino} with a pairing pseudogap \cite{randeria} above
$T_c$ for $U<0$.  Finally, both these models are realizable in cold atom
systems where, e.g., a Feshbach resonance can be used to change the
sign of the interaction. 

Our main results are: (1) we determine the characteristic temperature
scales for spin, charge and pairing correlations and the
superfluid $T_c$ as functions of the interaction
strength.  (2) We determine constant entropy contours in the
temperature-interaction plane, that give direct information on how
much the system has to be cooled, i.e., how low the entropy should be in
the experiments, in order to see interesting physics.  (3) We comment on
the difficulty of using adiabatic cooling in 2D.  (4) Finally we point
out the temperature range in which a simple observable like double
occupancy can be used for thermometry.  

The single band Hubbard Hamiltonian is
\begin{eqnarray}
\mathcal{H} &=& - t \sum_{(i,j),\sigma}
(c^\dagger_{i\sigma} c_{j \sigma}^{\phantom\dagger} + 
c^\dagger_{j\sigma} c_{i \sigma}^{\phantom\dagger} )
\nonumber 
\\
&+& U\sum_i (n_{i\uparrow}-\frac{1}{2})(n_{i\downarrow}-\frac{1}{2})
-\mu \sum_i n_i
\label{hamil}
\end{eqnarray}
where $c_{i \sigma}$ is the fermion destruction operator at site $i$
with spin $\sigma$, $n_{i\sigma}=c^\dagger_{i\sigma}c_{i\sigma}$ is the
density of fermions with spin $\sigma$, and $n_i=\sum_\sigma
n_{i\sigma}$.  For cold atoms $\sigma = \uparrow,\downarrow$ labels two
hyperfine states.   We consider near-neighbor hopping on a square
lattice with the kinetic energy $\epsilon( {\mathbf{k}} ) = -2t
(\cos{k_x a}+ \cos{k_y a})$.  In the determinantal QMC calculation, we
work in the grand canonical ensemble and tune the chemical potential
$\mu$ to obtain the desired density $\rho = \sum_i\langle n_i \rangle/N$
on an $N$-site lattice.  The parameters $t$ and $U$ can be directly
related \cite{zoller} to the lattice depth $V_0$, tuned by the laser
intensity, and to the interatomic interaction tuned by a Feshbach
resonance. We work in a parameter regime where only a single band
is populated in the optical lattice.  All energies and temperatures are measured in units of $t$ and distances in units of $a$.

\begin{figure}[t]
\includegraphics[width=2.5in,angle=0]{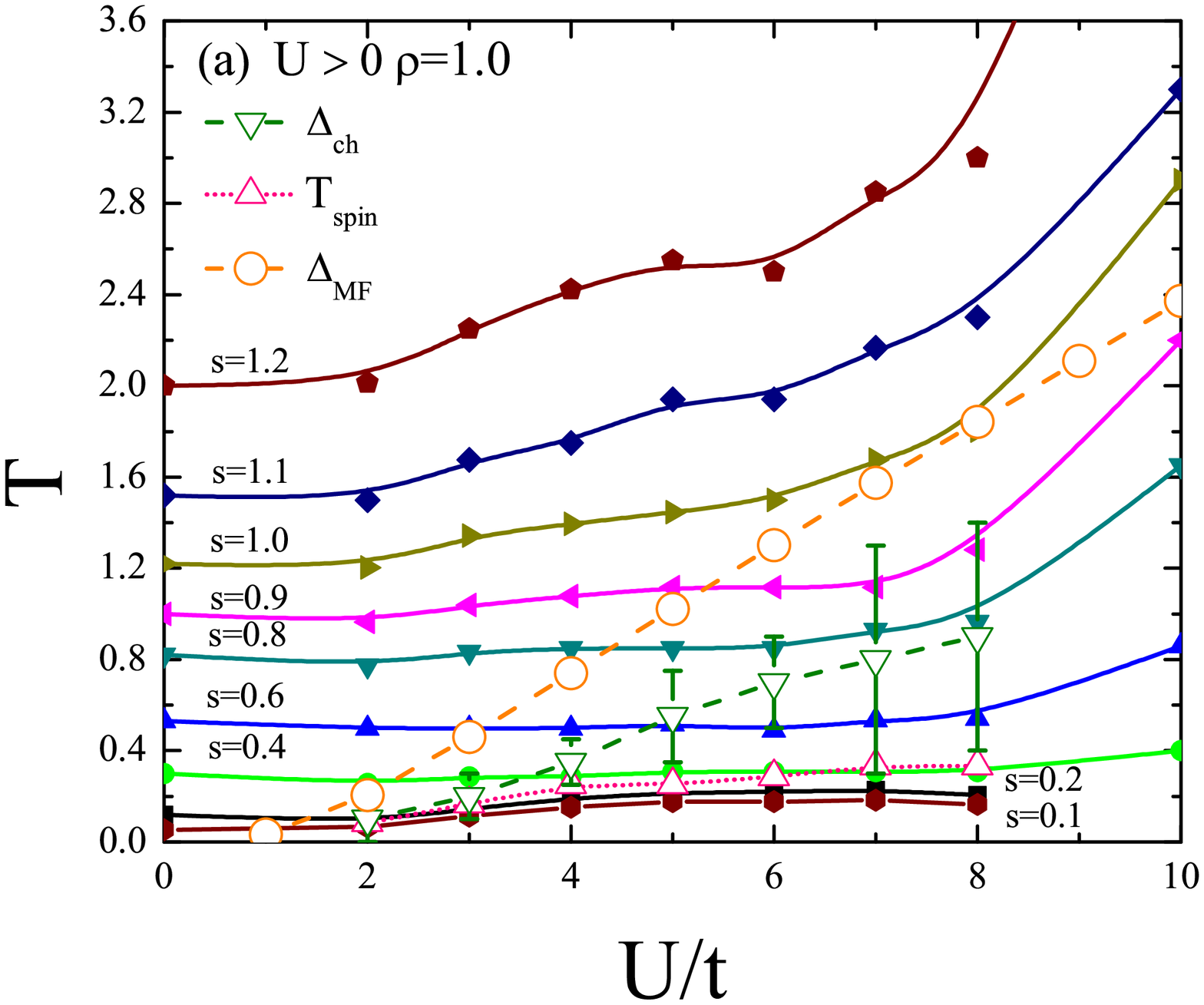}
\includegraphics[width=2.5in,angle=0]{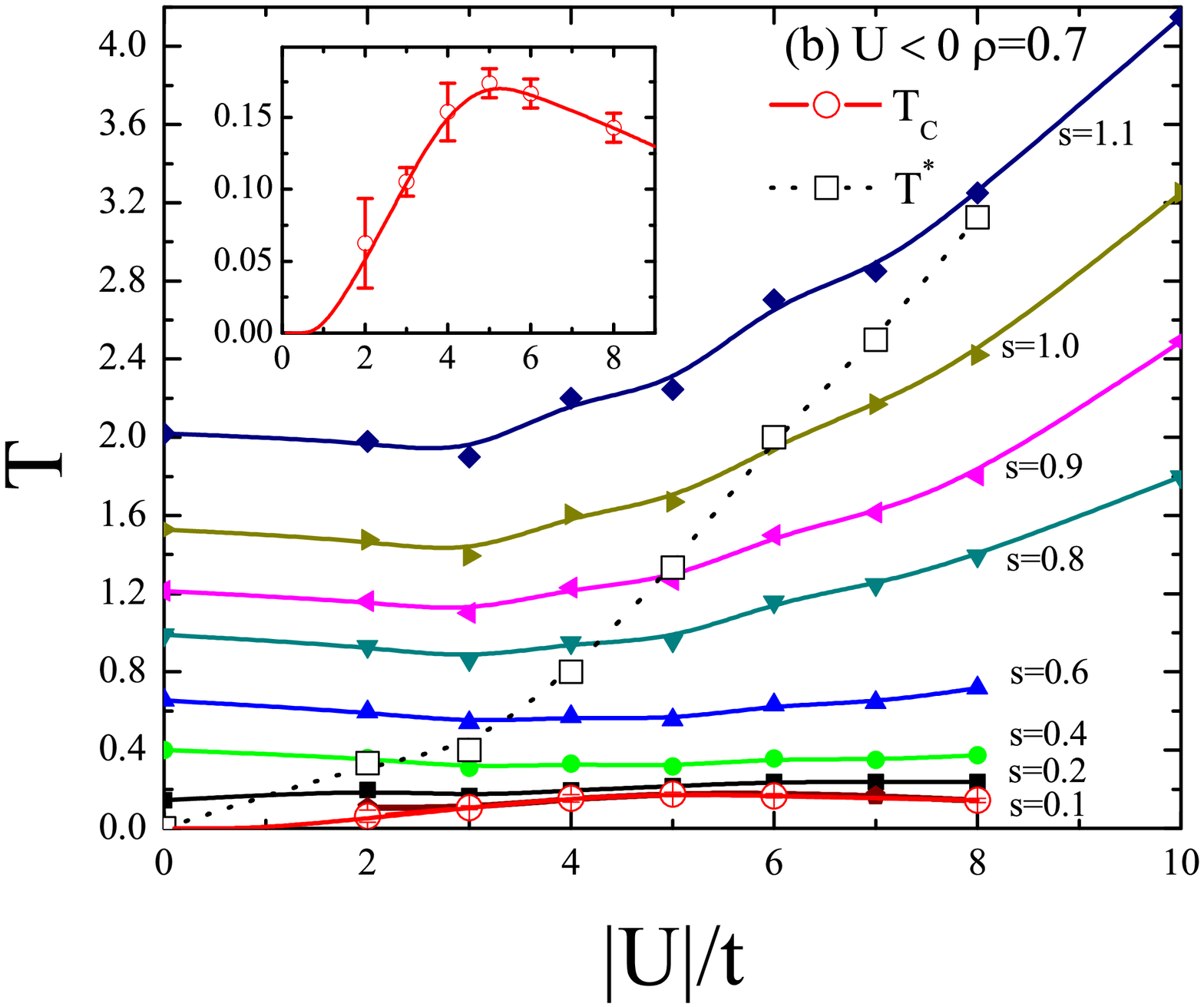}
\vspace{-0.2cm}
\caption{
Phase diagrams of the 2D Hubbard Models with (a) repulsive ($U > 0$)
interactions at half filling and (b) attractive ($U < 0$) interactions
away from half-filling, showing curves of constant entropy (per site) $s
= S/Nk_B$, obtained from QMC simulations on $N=10^2$ lattices.  In (a)
$\Delta_{\rm ch}$ is the scale for the formation of a charge gap,
$\Delta_{\rm MF}$ is the mean field charge gap (see text), and $T_{\rm
spin}$ is obtained from the peak in $\chi(T)$ in Fig.~2(b).  In (b), the
pairing pseudogap temperature $T^*$ is obtained from the peak in
$\chi(T)$ in Fig.~2(c), and the Berezinskii-Kosterlitz-Thouless transition $T_c$ is
estimated from the superfluid density, Fig.~\ref{fig2}(d); $T_c$ as a function of $U$ also shown in the inset. 
\vspace{-0.5cm}
}
\label{phasediagram}
\end{figure}

The results for the characteristic temperature scales (open symbols) and
constant entropy curves (filled symbols) in the $(T,U)$-plane are shown
in Fig.~\ref{phasediagram}.  We first describe the results for the
repulsive ($U>0$) model at half-filling $\rho = 1$ and then turn to the
attractive ($U<0$) case at arbitrary filling $\rho \ne 1$. 

\begin{figure*}
\includegraphics[width=1.72in,angle=0]{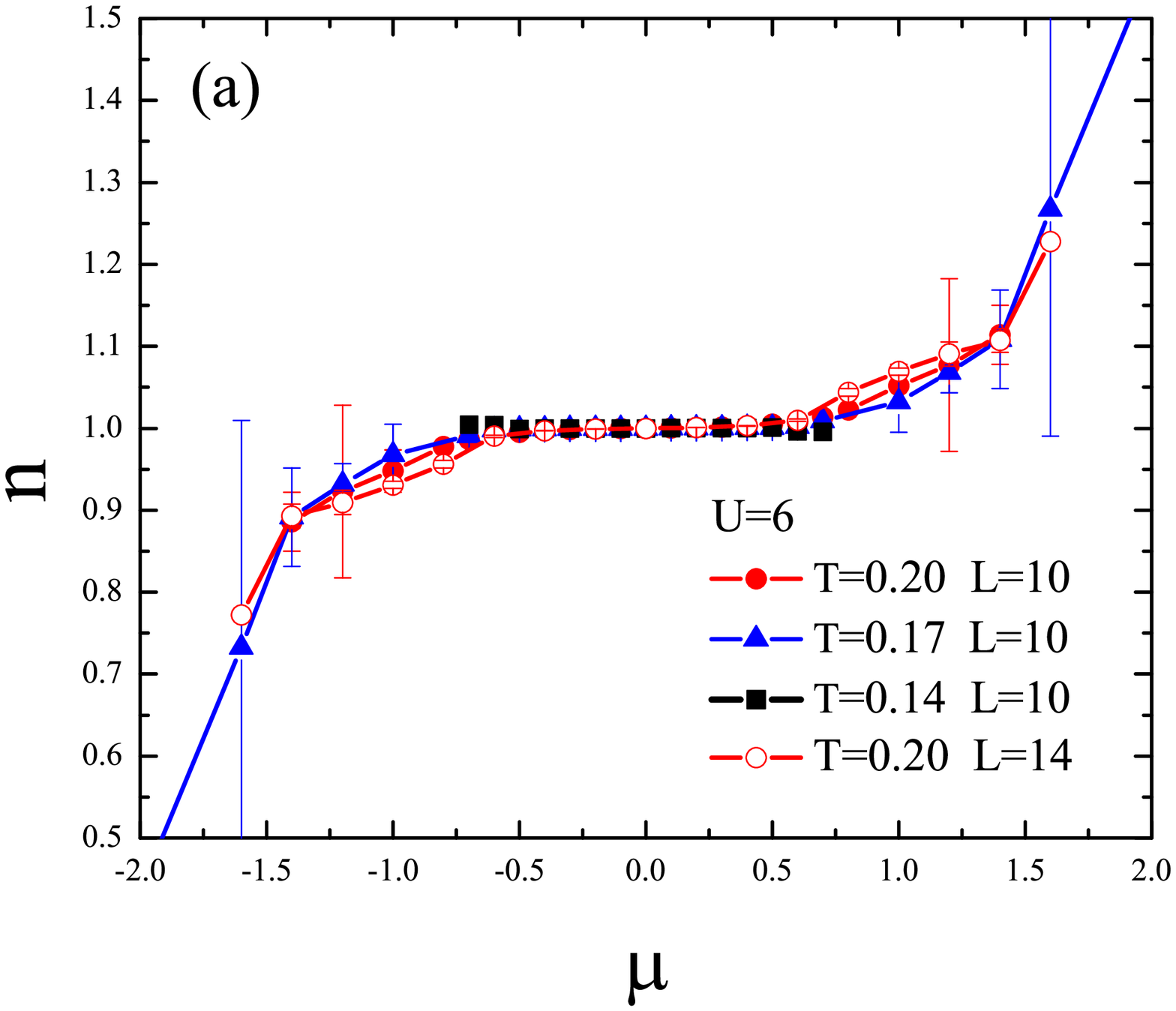}
\includegraphics[width=1.72in,angle=0]{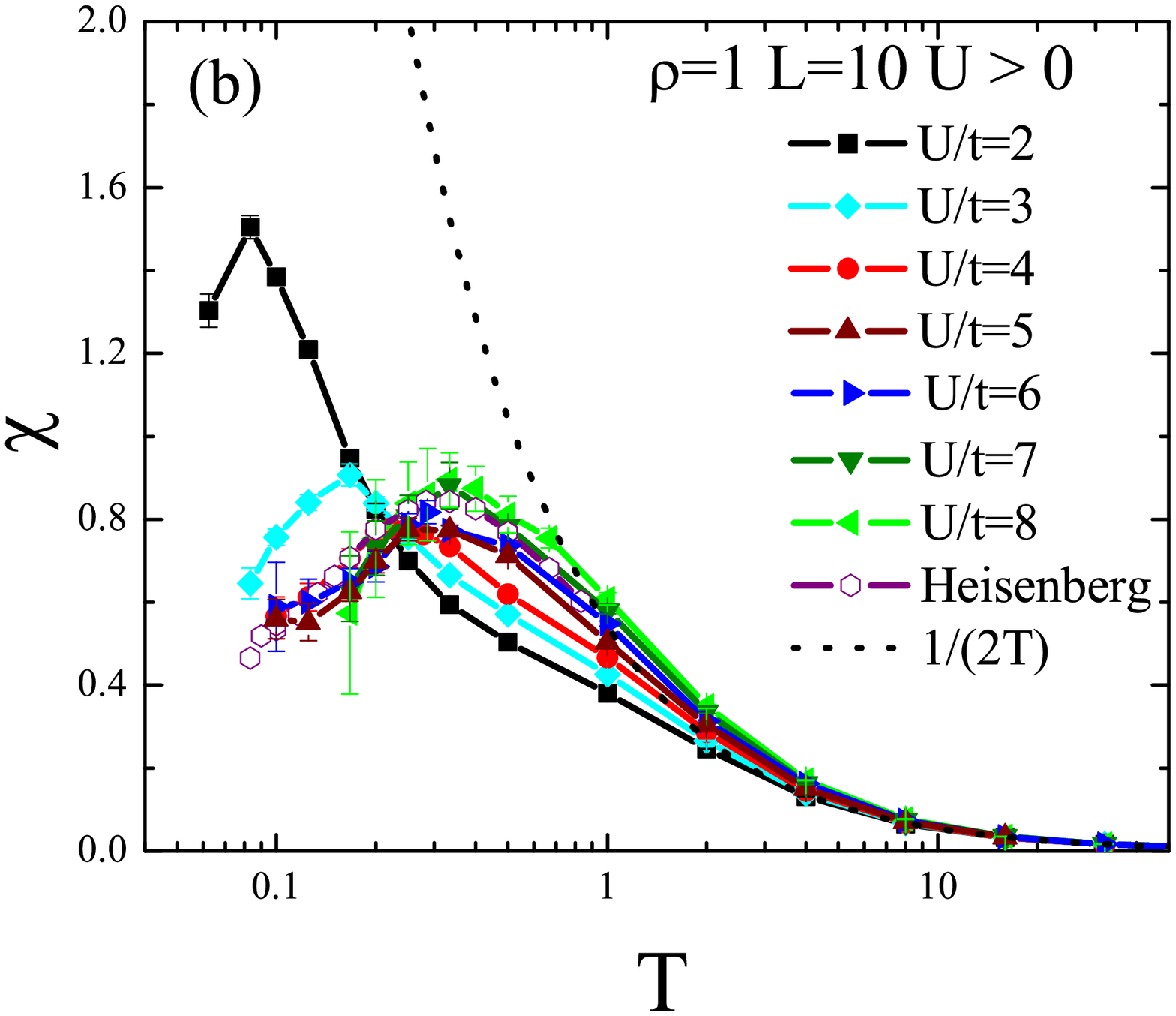}
\includegraphics[width=1.72in,angle=0]{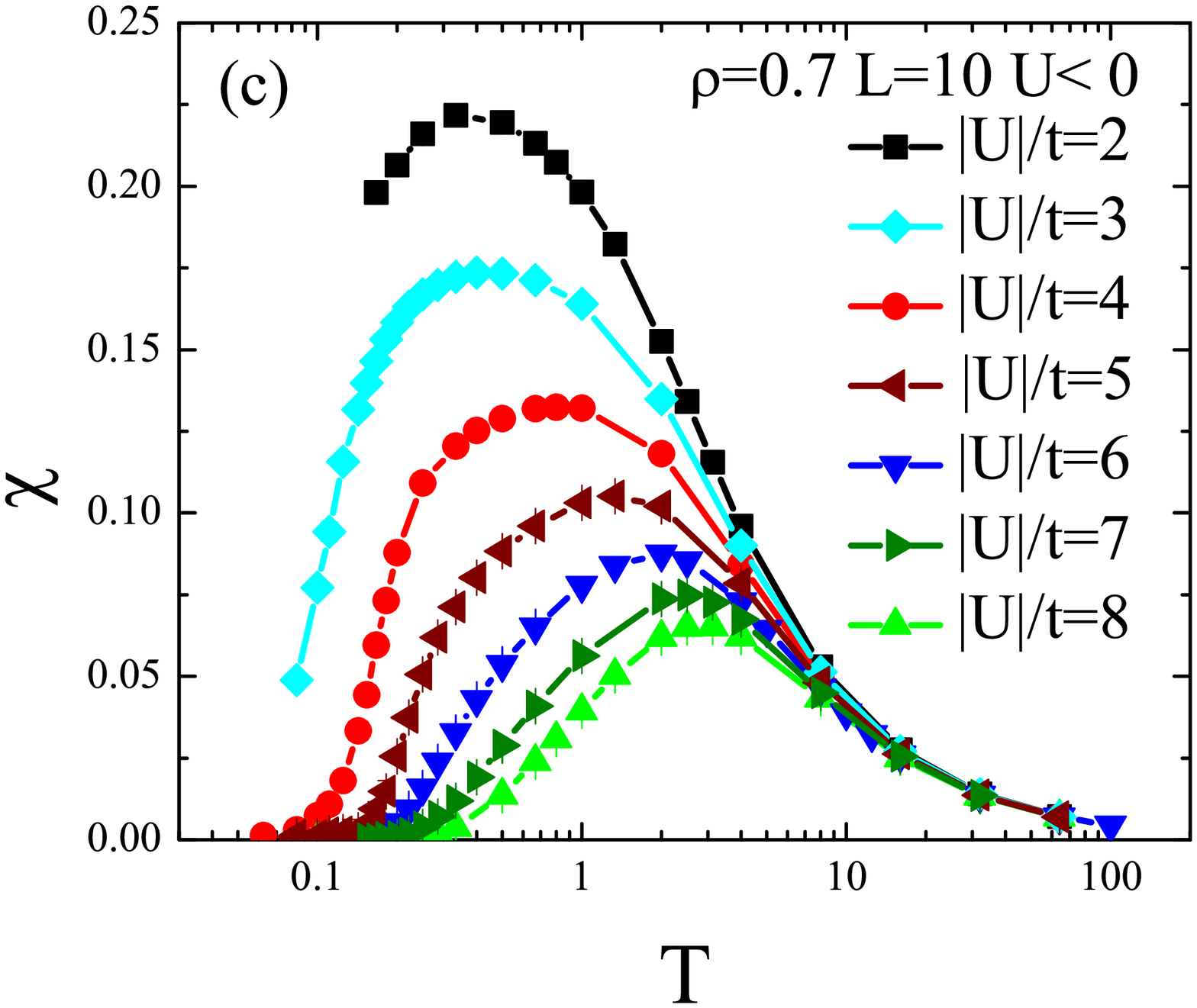}
\includegraphics[width=1.72in,angle=0]{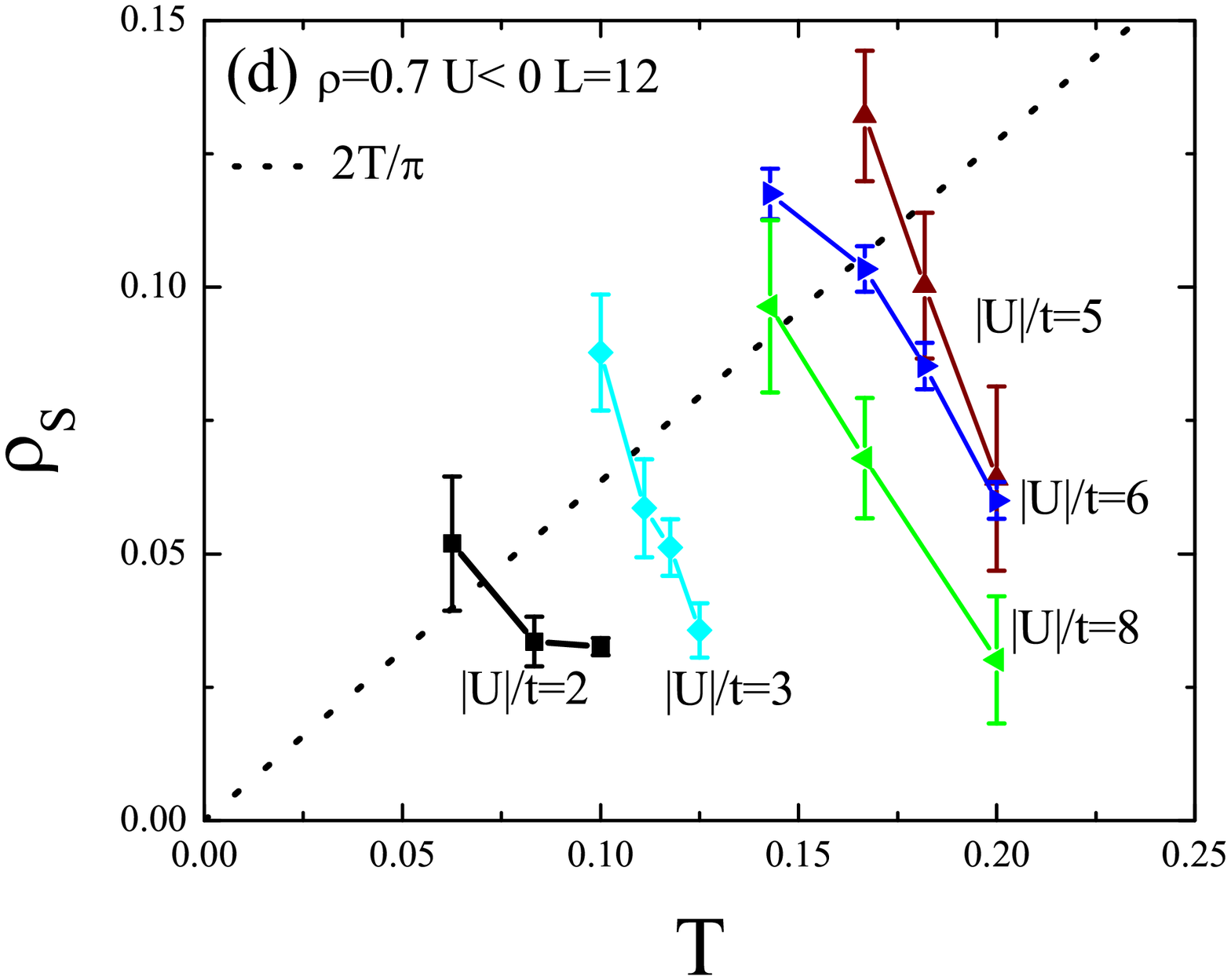}

\vspace{-0.3cm}
\caption{
(a) Density $\rho$ as a function of chemical potential $\mu$ for $U=6t$
and various $T$. Note that $\rho$ is pinned at unity for $|\mu| \le
\Delta_{\rm ch}$, the charge gap.  The large error bars away from
half-filling are a result of the sign problem.  Spin susceptibility
$\chi(T)$ for (b) positive $U$ at half filling and (c) negative $U$ at
density $\rho =0.7$. The peak in $\chi(T)$ is used to determine
characteristic temperature scales for the two models described in the
text. 
(d) The superfluid density $\rho_s(T)$ for various $U<0$ at $\rho
=0.7$. This is used to estimate the Kosterlitz-Thouless $T_c$ in 2D
using $\rho_s(T_{c})=({2/\pi})T$. 
\vspace{-0.5cm}
} 
\label{fig2}
\end{figure*}

\begin{figure}[t]
\includegraphics[width=2.5in,angle=0]{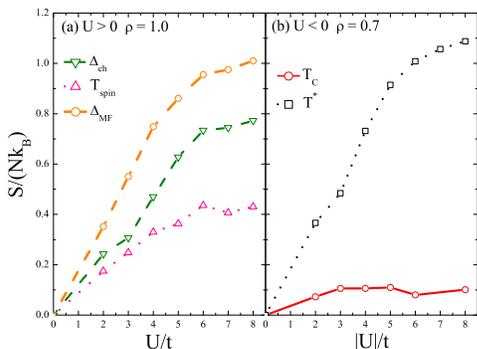}
\vspace{-0.5cm}
\caption{
Entropy per site $s= S/Nk_B$ as a function of $U$ at a temperature
corresponding to the characteristic scale for that $U$. (a) For $U>0$,
the entropy at $\Delta_{\rm ch}$, the charge scale, and $T_{\rm spin}$,
the spin scale is plotted.  (b) For $U<0$ the entropy at the pairing
scale $T=T^*$ and the superfluid  $T=T_c$ is plotted. 
\vspace{-0.75cm}
} 
\label{entropy} 
\end{figure}

\begin{figure}[t]
\includegraphics[width=3.0in,angle=0]{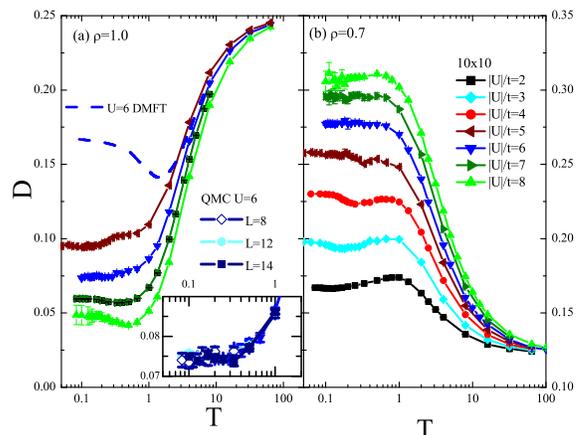}
\vspace{-0.5cm}
\caption{
Double occupancy $\langle n_{i\uparrow} n_{i\downarrow}\rangle$ as a
function of temperature $T$ for (a) $U>0$ and (b) $U<0$.  The legend in (b) 
also apply to the positive $U$ values in (a). Also shown are
the comparisons with dynamical mean field theory DMFT in panel (a) that
shows a much more pronounced anomalous region with $dD/dT < 0$. The
inset in (a) shows the lack of system size dependence in the $D(T)$
results.  
\vspace{-0.5cm}
} 
\label{doubleoccupancy} 
\end{figure}

%\noindent
{\bf Repulsive Hubbard model at half-filling:} 
Coming down from high temperatures the first scale encountered is the
energy gap to charge excitations. A lower scale corresponds to the
development of spin correlations.  The Mermin-Wagner theorem precludes a
finite temperature phase transition in this 2D system and both these
scales are crossovers.  Nevertheless, these scales (summarized in
Fig.~\ref{phasediagram}) have very clear signatures in physical
observables (Figs.~\ref{fig2},\ref{entropy},\ref{doubleoccupancy}).

\noindent \underline{Charge gap}:
The charge gap is seen in Fig.~\ref{fig2}(a) where we plot the density
$\rho$ as a function of the chemical potential $\mu$. Despite large
error bars, arising from the sign problem \emph{away} from half-filling,
we clearly see that the compressibility $d\rho/d\mu$ is very small; it
should vanish at $T=0$. The region $|\mu| < \Delta_{\rm ch}$ for which
$d\rho/d\mu \approx 0$ gives an estimate of the charge gap $\Delta_{\rm
ch}$.  The $U/t$-dependence of the charge gap is shown in
Fig.~\ref{phasediagram}(a): 
%we can define the temperature $T_{\rm ch} =
%\Delta_{\rm ch}$ such that 
for $T \gg \Delta_{\rm ch}$ we expect the system
to look effectively gapless while for $T \ll \Delta_{\rm ch}$ there is a gap
to fermionic excitations. 

For large $U$ the charge gap scales linearly with $U$ and is called the
Mott gap, but it becomes (exponentially) small for low $U$. To
understand this $U$-dependence we use a simple $T=0$ mean field theory
(MFT) of spin density wave (SDW) ordering that shows the system is
an insulator at $\rho = 1$ for any $U > 0$.  The $U=0$ metal is unstable
due to Fermi surface nesting, and the ground state is an insulator whose
gap is obtained from $N^{-1} \sum_{\bf k}[\epsilon_k^2 + \Delta_{\rm
MF}^2]^{-1/2} = {1/U}$.  The mean field charge gap is 
$\Delta_{\rm MF}\sim t \exp {-2\pi\sqrt{t/U}}$ for $U/t \ll 1$ and
smoothly crosses over to the Mott gap $\Delta_{\rm MF} = U/2$ for
$U/t\gg 1$.  This evolution \cite{schrieffer} from a SDW insulator to a
Mott insulator is the $U>0$ analog of the BCS to BEC crossover in
attractive Fermi systems.  We see from Fig.~\ref{phasediagram}(a) that
the $U/t$-dependence of $\Delta_{\rm MF}$ obtained from $T=0$ MFT overestimates 
the charge gap obtained from the
$\rho(\mu)$ analysis for moderate to large coupling.
The MFT is also quite misleading about the
finite temperature antiferromagnetic (AF) long range order, that is
necessarily absent in 2D. 

\noindent \underline{Spin correlations}: 
To study magnetic correlations we look at the uniform susceptibility
$\chi(T)$ for various $U/t$ shown in Fig.~\ref{fig2}(b). We find a
broad peak in $\chi(T)$ at a temperature $T_{\rm spin}$ below which
short range AF spin correlations are important.  We have also analyzed
the short range spin-spin correlation function $\langle S_i \cdot
S_{i+\delta}\rangle$ as a function of $T$ for various $U/t$ and found
that it too shows a growth in correlations at the temperature scale
$T_{\rm spin}$. 

To get a better feel for $T_{\rm spin}$ we also plot in
Fig.~\ref{fig2}(b) the uniform susceptibility for the 2D
nearest-neighbor $S=1/2$ Heisenberg AF \cite{makivic}, that describes
the low energy spin physics of the Hubbard model for $U/t \gg 1$. Here
we see a peak in $\chi$ at $T_{\rm spin} = J_{\rm AF}$.  To plot this
data together with our Hubbard model results, we have chosen (somewhat
arbitrarily) $U/t = 12$, so that the AF superexchange $J_{\rm AF} =
4t^2/U = t/3$. 

Except in weak coupling $U/t \ll 1$, where the charge and spin scales
are essentially identical, we find in Fig.~\ref{phasediagram}(a) that
the two scales are quite different for $U/t > 1$. In the large $U$
(Mott) limit we expect $\Delta_{\rm ch} = U/2 \gg T_{\rm spin} \simeq J_{\rm
AF} = 4t^2/U$.  For any $U/t$ we expect to see local moments below the
charge gap $T < \Delta_{\rm ch}$, and the build up of AF spin correlations
between the moments for $T < T_{\rm spin}$. 

\noindent\underline{Entropy}:
In cold atom experiments the entropy $S$ can be monitored more easily
than the temperature.  We calculate $S(T)$ from QMC in two
different ways. The first method \cite{binder,tremblay} is to integrate
down from infinite temperatures: $S(\beta) = \rho N \ln 4 + \beta
E(\beta) - \int_0^{\beta}d\beta^\prime E(\beta^\prime)$, where $\beta =
1/T$ and $E$ is the energy.  The second method \cite{paiva} is to
integrate up from $T=0$. Here we fit $E(T)$ data to a suitable
functional form, find the specific heat $C(T)=dE(T)/dT$ and calculate
$S(T)=\int_0^T C(T^\prime)/T^\prime$.  The methods agree to within a few
percent.

The curves of constant $S$ are plotted in the $(U,T)$-plane in
Fig.~\ref{phasediagram}(a). At weak coupling the charge and spin scales
are both exponentially small and the system looks like a highly
degenerate normal Fermi gas with a low entropy. It is only at very low
entropy (per particle) $s \ll \ln 4 \simeq 1.386$, that one observes
either the charge gap or build up of spin correlations for $U/t \le 1$.
For moderate coupling, say, $U/t \simeq 6$ we see that $s = \ln 2 \simeq
0.693$ already puts us below the charge gap scale. By lowering the entropy 
to $s\sim 0.4$ it is already possible to cross $T_{\rm spin}$. The difference 
between the entropy required to access the charge and spin scales grows as $U$ increases.
%However one needs to
%be at significantly lower $s \sim 0.1$, an order-of-magnitude below $\ln
%4$, to cross $T_{\rm spin}$.  
We note that $T_{\rm spin}$ sets the
maximum scale for the AF $T_{\rm Neel}$ if we were to couple 2D layers
to cut off the fluctuations and stabilize AF order in layered system
like the parent insulator of the high Tc superconductor. In
Fig.~\ref{entropy}, we summarize our results for the entropy per site
$S/Nk_B$ at the charge and spin temperature scales plotted as function
of $U$.

\noindent\underline{Cooling and Thermometry}: 
We next turn to a beautiful suggestion  of Werner {\it et.~al.}
\cite{werner} for adiabatic cooling in an optical lattice, that
exploits the anomalous $T$-dependence of the double occupancy $D =
\langle n_{i\uparrow} n_{i\downarrow}\rangle$ observed in dynamical mean
field theory (DMFT).  In brief, their idea is as follows. The entropy
and $D$ are related to the free energy $F$ via $S=-\partial
F/\partial T$ and $D = \partial F/\partial U$.  Using a Maxwell relation
we find $\partial S/\partial U = -\partial D/\partial T$.  For $U>0$,
the ``natural'' expectation is $dD/dT >0$ so that the temperature
increases along a constant $S$ curve upon increasing $U/t$. Thus the
DMFT observation of a significant anomalous region with $dD/dT <0$
implies that we can follow a curve of constant $S$ and cool the system
as the lattice is turned on, i.e., $U/t$ is increased.

We show in Fig.~\ref{doubleoccupancy}(a) the DMFT curve for $D(T)$ at
$U=6t$ contrasted with 2D QMC results.  The absence of a significant
anomalous regime with $dD/dT<0$ in 2D is likely a result of short range
spin correlations that are important in low dimensional systems but
neglected in DMFT.  We see that the constant-$S$ curves in
Fig.~\ref{phasediagram}(a) do not show a significant negative slope and
thus one cannot obtain adiabatic cooling in 2D.  

From Fig.~\ref{doubleoccupancy}(a) we see that the double occupancy
$D(T)$ is quite weakly $T$-dependent for $T < t$, but its monotonic
$T$-dependence at higher temperature suggests that $D(T)$, that can be
measured in cold atom experiments \cite{jordens}, can be used for
thermometry in the range $1 < T/t < 10$.  Such a thermometer would need
QMC results for $D(T)$ callibration. We show in the insert to
Fig.~\ref{doubleoccupancy}(a) that the double-occupancy, that is a
local observable, has no significant system size dependence and can
indeed be very accurately determined by QMC. 

%\noindent
{\bf Attractive Hubbard model at arbitrary filling:} 
The $U>0$ model away from $\rho = 1$ has a fermion sign problem in QMC,
so we turn to a different model to discuss superfluidity in a 2D lattice
system.  We change the sign of the interaction and examine the
attractive Hubbard model that exhibits s-wave pairing
\cite{moreo-scalapino} and a BCS to BEC crossover \cite{randeria} as a
function of $|U|/t$.  We will consider the case of $\rho = 0.7$ filling
for concreteness; we choose to work away from half-filling,  because at
$\rho = 1$ the results would be identical to those discussed above using
a well-known particle-hole transformation.

\noindent \underline{Pairing (Pseudogap) scale}:
The analog of the charge scale below which moments form in the repulsive
model, is the pairing scale $T^*$ in the attractive Hubbard model below
which double occupancy grows. Since pairing represents the onset of
singlet correlations, one of the simplest ways of probing this scale is
through the susceptibility $\chi(T)$ plotted in Fig.~\ref{fig2}(c) for
a sequence of attractive couplings $|U|/t$. The peak in $\chi$ is a
measure of the pairing scale, for below it the spin response is strongly
suppressed by the formation of pairs.  The $|U|/t$-dependence of $T^*$
as determined from $\chi$ is plotted in Fig.~\ref{phasediagram}(b).
$T^*$ essentially has the same $|U|$-dependence as the $T=0$ pairing
gap: exponentially small in $|U|/t$ for the weak coupling BCS limit and
proportional to $U$ in the strong coupling Bose limit. For $T < T^*$
there is a pseudogap \cite{randeria} due to pairing correlations in the
density of states, even above $T_c$ where the system is not superfluid. 

\noindent \underline{Berezinskii-Kosterlitz-Thouless Transition}:
At lower temperatures there is a transition to a superfluid phase
in the attractive Hubbard model for $\rho \ne 1$.  There is
algebraic order in the pair-pair correlation function
\cite{moreo-scalapino} below $T_c$ and a non-zero superfluid density
$\rho_s$. We calculate $\rho_s$ from the transverse current-current
correlation function \cite{scalapino-white-zhang} and use the universal
jump in $\rho_s(T_c^{-}) = 2T_c/\pi$ to estimate the BKT $T_c$ as shown
in Fig.~\ref{fig2}(d). At a filling of $\rho=0.7$ there is a maximum
in $T_c \simeq 0.2 t$ as a function of attraction at $|U|/t \simeq 5$ as
seen from Fig.~\ref{fig2}(d). The non-monotonic dependence of $T_c$ on
$|U|/t$ is expected with an exponentially small $T_c$ in the weak
coupling BCS limit and $T_c \sim t^2/|U|$ in the strong coupling BEC
limit. 

\noindent\underline{Entropy, Cooling and Thermometry:}
We next calculate the entropy using the methodology described above and
plot the curves of constant $S$ in the $(|U|,T)$-plane in
Fig.~\ref{phasediagram}(b). We see from Figs.~\ref{phasediagram}(b)
and 3(b) that for a coupling $|U|/t \simeq 5$ at which the $T_c$ peaks
for $\rho = 0.7$, one only needs to lower the entropy $S/Nk_B < 0.8$ to
enter the pseudogap regime below $T^*$. To actually observe
superfluidity one needs to cool below the BKT transition, that
corresponds to $S/Nk_B < 0.1$.

The double occupancy $D(T)$ for $U<0$ in
Fig.~\ref{doubleoccupancy}(b) shows a rather small regime
of anomalous behavior, which now corresponds to $dD/dT >0$.  Thus the
prospects of using adiabatic cooling \cite{werner} in 2D do not look
promising for the attractive case either. However,
Fig.~\ref{doubleoccupancy}(b) does suggest that $D(T)$ can be used as
an effective thermometer for the high temperature range $t < T < 10t$.

{\bf Conclusions:}
Current fermion optical lattice experiments \cite{schneider} have achieved an entropy per
particle $\simeq \ln 2$, sufficient to observe the
charge gap in the repulsive Hubbard model or the pairing pseudogap in
the attractive case. Observing antiferromagnetic correlations or
superfluidity in 2D systems will require a further reduction in the
entropy by a factor of three or more.  It is possible that the inhomogeneous density in a
trap can lead to a redistribution of entropy with some regions having a much lower entropy than
others.  
%% We plan to investigate the effect of the trap within the local
%% density approximation for $U<0$ where uniform systems can be studied at
%% arbitrary $\mu$ via QMC without any sign problem. 

We acknowledge support from the Brazilian agencies CNPq and FAPERJ (TP), 
ARO Award W911NF0710576 with funds from DARPA OLE Program (RTS), 
ARO W911NF-08-1-0338 (MR and NT), NSF-DMR 0706203 (MR), and the use of computational facilities at the
Ohio Supercomputer Center.

\vfill\eject


\vspace{-0.5cm}

\begin{thebibliography}{99}

\vspace{-0.5cm}

\bibitem{esslinger}
M.~Kohl {\it et al.},
%% H. Moritz, T. Stoferle, K. Gunter, and T. Esslinger
Phys.~Rev.~Lett.~{\bf 94}, 080403 (2005).

\bibitem{ketterle}
J.K.~Chin {\it et al.}, 
%% D. E. Miller, Y. Liu, C. Stan, W. Setiawan, C. Sanner, 
%% K. Xu  and  W. Ketterle,
Nature {\bf 443}, 961 (2006).

\bibitem{jordens}
R.~Jordens {\it et al.}, 
%% N. Strohmaier, K. Gunter, H. Moritz, and T. Esslinger,
Nature {\bf{455}}, 204 (2008).

\bibitem{schneider}
U.~Schneider {\it et al.}, 
%% L. Hackermuller, S. Will, Th. Best, I. Bloch, 
%% T.A. Costi, R.W. Helmes, D. Rasch and A. Rosch, 
Science {\bf{322}}, 1520 (2008).

\bibitem{hofstetter}
W.~Hofstetter {\it et al.}, 
%% J. I. Cirac, P. Zoller, E. Demler, and M. D. Lukin,
Phys.~Rev.~Lett.~{\bf 89}, 220407 (2002).

\bibitem{zoller}
D. Jaksch and P. Zoller,
Ann.~Phys.~{\bf 315}, 52 (2005);
D.~Jaksch {\it etal.}, 
%% C. Bruder, J. I. Cirac, C. W. Gardiner, and P. Zoller
Phys.~Rev.~Lett.~{\bf 81}, 3108 (1998).

\bibitem{scalapino}
D.J.~Scalapino in ``Handbook of High Temperature Superconductivity'',
edited by J.R. Schrieffer and J.S.~Brooks, (Springer, 2007);
arXiv:cond-mat/0610710.

\bibitem{andersonAtoZ}
P.W.~Anderson {\it et al.}, 
%% P A Lee, M Randeria, T M Rice, N Trivedi and F C Zhang,
J.~Phys.~Cond.~Mat.~{\bf 16}, R755 (2004). 

\bibitem{giamarchi}
T.~Giamarchi, ``Quantum physics in one dimension'' (Oxford, 2004).

\bibitem{georges}
A.~Georges {\it et al.},
%% G. Kotliar, W. Krauth, and M. J. Rozenberg,
Rev.~Mod.~Phys.~{\bf 68}, 13 (1996).

%\bibitem{lewenstein}
%M. Lewenstein, A. Sanpera, V. Ahufinger, B. Damski, A. Sen De, U. Sen,
%Adv. Phys. {\bf 56} 243 (2007). 

\bibitem{hirsch}
J.E.~Hirsch, Phys.~Rev.~B {\bf 28}, 4059 (1983).

\bibitem{moreo-scalapino}
A.~Moreo and D.J.~Scalapino,
Phys.~Rev.~Lett.~{\bf 66}, 946 (1991).

\bibitem{randeria}
M.~Randeria {\it et al.}, 
%% N. Trivedi, A. Moreo, and R. T. Scalettar,
Phys. Rev. Lett. {\bf 69}, 2001 (1992);
N.~Trivedi and M.~Randeria,
Phys.~Rev.~Lett.~{\bf 75}, 312 (1995).

\bibitem{schrieffer}
J.R.~Schrieffer, X.G.~Wen, and S.C.~Zhang
Phys.~Rev.~B {\bf 39}, 11663 (1989).

\bibitem{makivic}
M.S.~Makivic and H-Q.~Ding, Phys.~Rev.~B {\bf 43}, 3562 (1991).

\bibitem{binder}
K.~Binder, Z.~Phys.~{\bf 45}, 61 (1981).

\bibitem{tremblay}
A.-M.~Dare {\it et al.}, 
%% L. Raymond, G. Albinet, and A.-M.~S. Tremblay, 
Phys.~Rev.~B {\bf 76}, 064402 (2007).

\bibitem{paiva}
T.~Paiva {\it et al.}, 
%% R. T. Scalettar, C. Huscroft, and A. K. McMahan,
Phys.~Rev.~B {\bf 63}, 125116 (2001).	

\bibitem{werner}
F.~Werner {\it et al.}, 
%% O. Parcollet, A. Georges, and S.R.~ Hassan, 
Phys.~Rev.~Lett.~{\bf 95}, 056401 (2005).

\bibitem{scalapino-white-zhang}
D.J.~Scalapino, S.R.~White, and S.C.~Zhang,
Phys.~Rev.~B {\bf 47}, 7995 (1993).

%\bibitem{leo}
%L. De Leo {\it et al.}, 
%% C. Kollath, A. Georges, M. Ferrero, and O. Parcollet, 
%arXiv:0807.0790.
%\textcolor{red}{[need this?]}

%\bibitem{koetsier}
%A. Koetsier {\it et al.}, 
%% R.~A. Duine, I. Bloch, and H.~T.~C. Stoof, 
%Phys.~Rev.~B {\bf 77} 023623 (2008).
%\textcolor{red}{[?]}

%\bibitem{snoek}
%M. Snoek {\it et al.}, 
%% I. Titvinidze, C. Toke, K. Byczuk, and W. Hofstetter, 
%New J.~Phys.~{\bf 10}, 093008 (2008).
%\textcolor{red}{[?]}


\end{thebibliography}
\end{document}